\documentstyle [11pt] {article} \textwidth 170mm\textheight
238mm \topmargin - 20mm \oddsidemargin -5mm\evensidemargin -5mm
\begin{document} \footskip=1000mm

\begin {center} {\Large Lebedev Physical Institute RAS, Moscow, Russia}
\end {center}
\begin {center} {\Large Branch of Nuclear Physics and Astrophysics} \end {center}
\begin {center} {\Large Department of High Energy Physics} \end {center}
\begin {center} -------------------------------------------- \end {center}
\vskip 4cm
\begin {center} {\large English version of the Preprint FIAN No 35, 1997
\\enlarged by Appendix, some comments and foot-notes} \end {center}
\vskip 2cm \begin {center} {\large E.G.Bessonov} \end {center}
\begin {center} {\Large \bf To Foundations of Classical Electrodynamics} \end {center}
\vskip 7cm \begin {center} {\large Moscow  1997} \end {center}
\newpage

\footskip=10mm
\setcounter {page} {3}

\newpage
              \begin{abstract}
In the present work a number of questions concerning foundations of the
classical electrodynamics are discussed. First of all these are the law
of conservation of energy  and the introduction of particles in
classical electrodynamics. We pay attention to a logical error which
appears at the interpretation of the Poynting's theorem ostensibly
following from the equations of Maxwell-Lorentz as the law of
conservation of energy. It was shown that the laws of
conservation of energy and momentum of the system of electromagnetic
field and charged particles does not follow from the equations of
electrodynamics. The violation of these laws is displayed when the
energy of particles is changed. Particular examples are considered
which make it possible to restrict a possible form of fields of a
non-electromagnetic origin. Thus we transfer the essence of
difficulties of electrodynamics deeper in its foundations. We hope
that this work will permit to produce a more comprehensive analysis
and to stimulate the further development of the foundations of the
classical and quantum electrodynamics.

Originally the main part of this paper was published as the Internal Report
FIAN in 1975, the Preprint FIAN No 196, 1980 and in the preprint FIAN No 35,
1997 (enlarged by a spectrum of opinions). This paper is the Preprint FIAN
No 35, 1997 enlarged by Appendix, some comments and foot-notes.
\end{abstract}
\newpage

\empty \vspace * {4cm}
{ 
                           \tableofcontents

\newpage

                       \section {Introduction}

Usually speaking about difficulties arising in classical electrodynamics,
for example, at introduction in it particles, establishment of the
equation of motion of particles in view of radiation reaction forth
and other it is emphasized that these difficulties are non-principle, that
classical electrodynamics is the consistent relativistic theory. In addition
everybody refer to the laws of conservation of energy, linear and angular
momentum for particles and fields.

In the present work we show that similar references on the conservation laws
are incorrect. We pay attention to a typical logical error in textbooks on
classical electrodynamics when the law of conservation of energy (Pointing's
theorem) is proved for the case of a system consisting of electromagnetic
field and charged particles. The violation of this law is displayed when the
energy of particles is changed. Private examples are considered which
allow to impose certain restrictions on a possible form of fields of
non-electromagnetic origin. Thus we transfer the essence of difficulties of
electrodynamics deeper in its foundations. 

\section {The laws of conservation in electrodynamics}

Below we shall present the typical proof of the law of conservation of
energy in electrodynamics of Maxwell-Lorentz for a system consisting of
an electromagnetic field and charged particles to specify then a logic
error. There are three basic moments in the proof:
\cite[p.106]{landau}-\cite{sokol}.

1. From Maxwell equations an equation follows
         \begin {equation} 
        {\partial \over \partial t} \int {|\vec E | ^2 + | \vec H |
         ^2\over 8\pi} dV +
         \int \vec j\vec E dV = 0,
                \end {equation}
where $\vec E$, $\vec H$ are vectors of an electric and magnetic field
strengths, $\vec j = \rho \vec v$ vector of a current density,
$\rho$ a particle charge density, $\vec v$ vector of velocity of a
motion of the given element of volume of a charge of the particle
and the integration is made over the whole space.

2. From a relativistic equation of motion of a charged particle and
definition of work the law of energy change of a particle is derived
            \begin {equation} 
            {d\varepsilon \over dt} = \int \rho \vec E\vec v dV.
                 \end {equation}

It is considered that the energy of a particle is the value
                       \begin {equation} 
                  \varepsilon = {mc^2 \over \sqrt {1 - v^2/c^2}},
                       \end {equation}
where $m$ is a mass of a particle, $v$ a velocity of a particle, $c$ a
velocity of light. The dependence of the particle's energy on velocity is
determined by the requirements of the special theory of relativity. At
introduction of particles in electrodynamics it is usually postulated,
that the energy of a particle consists partially on the energy of it's
own electromagnetic field connected with a particle $\varepsilon ^ {em}
$, and partially from the energy of it's field of a non-electromagnetic
origin $\varepsilon ^ {not \, em} $ ($\varepsilon = \varepsilon ^ {em}
+ \varepsilon ^ {not \, em} $).

3. It is postulated that the value
              \begin {equation} 
          W = {| \vec E | ^2 + | \vec H | ^2\over 8\pi}
              \end {equation}
is the density of the energy of an electromagnetic field.

4. By elimination of the value $\int \vec j \vec E dV =
\int \rho \vec v\vec E d V$ from the equations (1) and (2) we will
obtain the expression $(d/dt)\varepsilon _ {\Sigma} = 0$. It follows
from this equation that the value
          \begin {equation} 
         \varepsilon _ {\Sigma} = \int
         {| \vec E | ^2 + | \vec H | ^2\over 8\pi} dV + \sum _i
         \varepsilon _i = const, \end {equation}
where $\varepsilon _i$ is the energy of a particle with an index $i$,
$\varepsilon _ {\Sigma}$ the total energy of a system which is a sum of
energies of an electromagnetic field and particles in the hole space.
The expression (5) is treated as the law of conservation of energy in
electrodynamics.

Vectors $\vec E$, $\vec H$ in (5) represent a total electromagnetic
field created by a system of particles.
Volume integral of the energy density of the electromagnetic field
represents a system of a self-energy  (inertial energy) of particles,
an electromagnetic interaction energy of particles (potential energy)
and an energy of free fields. In a general case these components of
energy are not separated. The radiation field can be selected only in a
wave zone.

The error in the presented proof consist in unification of physically
inconsistent equations (1) - (4) in one system. This inconsistency
occurs because of non-consistent introduction of particles in
electrodynamics. Namely, if we start from the Maxwell equations and
postulate (4) then we should adopt that the energy of a charged
particle $\varepsilon _i$ in (5) partially consists of an energy of own
electromagnetic field created by it. But the energy of a particle is
included in the first term of this equation\footnote {\normalsize It
remains in a case of one uniformly moving particle.}. It means that the
energy of own electromagnetic field of a particle is contained in the
equation (5) twice. Just this fact leads to the logic error in the
proof of the law
of conservation of energy and because of which the equation (5)
conflicts with the initial equations. It is not difficult to understand
the reasons according to which this error was not exposed for a so
prolonged time.

1. Usually the relativistic expressions of energy, momentum and mass
of particles are introduced within the framework of the special theory
of relativity. From this theory follows that the energy and
momentum of particles should have certain transformation
properties regardless to a nature of an origin of these values. Mass
acts in a role of coefficient of proportionality between vectors
of a momentum and velocity of particles.

2. In the electrodynamics the received values of energy and momentum of
particles are introduced through the equations of a motion. The joint
solution of the equations of motion for fields and particles
reduces to the expression (5), which is treated as the law of
conservation of energy. At this it is postulated that the value (4) is a
density of the energy of an electromagnetic field.

3. The nature of mass, energy and momentum of particles are discussed
later after the electrical and magnetic fields created by a uniformly
moving spherically symmetrical particle and connected with these fields
the energy and momentum of the particle of the electromagnetic origin
were calculated. It turned out that these values have not correct
transformation properties following from the special theory of
relativity\footnote{\normalsize Besides a correct factor $1/\sqrt
{1-v^2/c^2} $ there are the incorrect factors $ (1 + v^2/3c^2) $ in
energy and (4/3) in the momentum of particles\cite {ivan}.}. For
this purpose and with the purpose of holding of a particle charge in
equilibrium it is necessary to introduce attraction fields of
non-electromagnetic origin the energy and momentum of which have wrong
transformation properties of the other form. It is postulated that in a
sum the fields of the electromagnetic and non-electromagnetic origin
are reduced to experimentally observable values of energy and momentum
of particles\footnote {In the case of a non-uniform motion the concepts
of energy and momentum of the electromagnetic origin were not
discussed.  Non-obviously they are taken equal to the appropriate
values for the particles moving uniformly with the same velocity.}.

If after these remarks we return to the expression (5) then it is
not difficult to notice that the energy of particles of an
electromagnetic origin is introduced in this expression twice, and
consequently the standard treatment of this expression is incorrect.

Unfortunately the laws of conservation, as a rule, were proved only
to emphasize the consistency of the electrodynamics its perfection and
in that form in which they were received they did not usually used
since on the basis of only the laws of conservation it is possible to
solve a small number of simple problems not representing
practical interest\footnote {\normalsize It is possible to point out
only on close questions connected, for example, with renormalization of
mass in a classical electrodynamics (see [1], $\S$ 65, Problem 1,
where in the expression for a sum of energies of particles and fields
created by them and written down with accuracy of up to the terms of
the second order of $v/c$ without the analysis of legitimacy the energy
of particles of an electromagnetic origin is rejected).}. Therefore
the error in the proof, which would be possible to establish by a
comparison of the solutions following from the laws of conservation and
from the equations of a motion, on particular examples was not detected.
The next example illustrate this conclusion.

\subsection {Example 1}

There is an immovable spherically symmetrical charge  $q$. On a distance
$a$ a particle with a charge $e$ and mass $m$ is located. In some moment
the particle start to move and, being accelerated, leaves to infinity.
Compare the energy of a particle calculated from the law of the energy
conservation (5) and from the equations of motion.

Let us write down the expression (5) for an initial and a final states
of the system and equate received expressions. Then we shall determine
the kinetic energy of a particle
              \begin {equation} 
      T = mc^2 (\gamma - 1) = {eq\over a} - \varepsilon ^ {em} _ {rad}
         - (\varepsilon ^ {em} _ {e} - \varepsilon ^ {em} _ {e0}), \end
       {equation}
where $\gamma = 1/\sqrt {1-v^2/c^2} $, $eq/a$ is the initial potential
energy of the particle, $\varepsilon _ {rad} ^ {em} $ the energy of an
electromagnetic field radiated by the particle, $\varepsilon _e^ {em} $
and $\varepsilon ^ {em} _ {e0} $ the energy of the own electromagnetic
field of the moving particle and the particle at rest accordingly.

Calculation of a kinetic energy of a particle by the solution of the
equation of a motion of the particle in a given field of the charge
$q$ will lead to an expression
             \begin {equation} 
             T = {eq\over a} - \varepsilon ^ {em} _ {rad}.  \end {equation}

As was to be expected, an extra term in (6) is equal to
a difference between own energies of electromagnetic fields of
accelerated and motionless particles. For the spherically symmetric
distribution of   particle's charge the value
$\varepsilon ^ {em} _ {e} - \varepsilon ^ {em} _ {e0} = \varepsilon ^
{em} _ {e0} [\gamma (1 + v^2/3c^2) - 1] $ \cite {ivan}.

\subsection {Example 2}

Investigate the transformation properties of the energy $\varepsilon _
{\Sigma} $ determined by the expression (5).

For the case of one uniformly moving particle the value $\varepsilon _
{\Sigma} $ depends on velocity by the law which differ from (3)
as it has not correct  transformation properties since this properties
has the second but has not the first term of expression (5) which
is equal to the electromagnetic self-energy of the particle \cite {ivan}.
The similar statement is valid for the beam of charged particles
where the situation is intensified in addition by the circumstance
that besides the electromagnetic self-energy of particles the first
term in (5) will include the electromagnetic energy of interactions of
particles which have incorrect transformation properties as well.

Thus transformation properties of the energy $\varepsilon _ {\Sigma} $
do not satisfy to the requirements of the special theory of relativity.

                       \subsection {Remarks}

1. In the case of pointlike particles the first term in (5) and
consequently the value  $\varepsilon _ {\Sigma} $ are infinite. They
are changed to the infinitely large value when the particle's
velocity change is finite. Since in the process of the evolution of a
system the velocity of particles can be simultaneously increased or
decreased then contrary to (5) in this case $\varepsilon _ {\Sigma} \ne
const$. On the other hand for the pointlike particles the proof of the
laws of conservation operating with indefinitely large values cannot
be correct.

It is interesting to know why did not anybody pay attention
to this moment till now?  For example, in the textbook~\cite[p.266]
{landau} the law of conservation of energy is proved for pointlike
particles, but in the process of the proof there is no scientific
substantiation of the correctness of the mathematical operations with
the indefinitely large values. At the same time later discussing the
difficulties which appear in the electrodynamics when the runaway
solutions appear we can read in this textbook: "A question can arise.
How the electrodynamics satisfying to the law of conservation of
energy can lead to an absurd result when a free particle increases its
energy by unlimited way. The roots of this difficulties are, actually,
in the mentioned earlier infinite electromagnetic "mass" of elementary
particles. When we write finite mass of a charge in the equations of
motion we, as a matter of fact, add to it formally an infinite negative
"mass" of non-electromagnetic origin which together with an
electromagnetic mass would result in final mass of a particle.  Since,
however, the subtraction of one from another of two infinities is not
quite correct mathematical operation then it lead to a number of
further difficulties including to the specified one here".

The logic however suggests that it is necessary the reasoning about
non-correctness of mathematical operations to transfer to the proof
of the law of conservation of energy instead of to base on this law.

2. In the electrodynamics of Maxwell-Lorentz there could not be a model
of particles with pure electromagnetic nature of mass. Differently all
energy would have electromagnetic nature and the second term in (5)
should be absent. On the other hand the energy of charged particle
$\varepsilon $ cannot have pure non-electromagnetic nature since in a
case, for example, of one particle the energy of extraneous forces
applied to a particle will be transformed not only to the value
$\varepsilon $ and to the energy of the emitted radiation but also to
the electromagnetic self-energy $\varepsilon ^ {em} $ of the particle.

From the equations of electrodynamics instead of (5) a conclusion about
conservation of a sum of energies of electromagnetic and
non-electromagnetic origin possessing the correct transformation
properties should follow. From these energies on certain unknown for
the present principles it could be possible to select the energy of
particles.

3. When the equations of motion and (2) are solved then only an
external and a part of an electromagnetic self-field of a particle
corresponding to a radiation reaction forth are taken into account.
At the same time the inertial field of the particle is
rejected\footnote {\normalsize In a non-relativistic case the electric
field strength corresponding to the radiation reaction forth is
proportional to the derivative of the particle acceleration and the
electric field strength corresponding to the inertial field is
proportional to the particle acceleration.} since it is considered
that the fields of the non-electromagnetic origin produce a field
of force equal to it by value but with the opposite sign\footnote
{\normalsize In case of the pointlike particles the inertial fields,
theirs energies and the electromagnetic masses are infinitely large
values.}. Such "volitional" renormalization of a mass in the equations
of motion of particles and also the rejection of the terms in the
individual examples similar to the last term in (6) contradicts to
a general principles which are guided at a derivation of (5).

In avoidance of misunderstanding we shall emphasize that the mentioned
inconsistency of the equations takes place only in the electrodynamics
of fields and particles and is displayed at the energy change of
particles. For example, in the case of macroscopic electrodynamics the
inconsistency of the introduced equations with the equations of the
electromagnetic field does not appear.

So if the material bodies are motionless then it is possible to write
down  the current density in them as
            \begin {equation} 
         \vec j = \lambda (\vec E + \vec E^ {extra}), \end {equation}
where $\lambda$ is a factor the electrical conductivity, $\vec E^
{extra} $ the vector of the electric field strength of the extraneous
forces.

In this case the expression (1) can be presented as
               \begin {equation} 
       {\partial \over \partial t} \int {| \vec E | ^2 + | \vec H |
   ^2\over 8\pi} dV = \int \vec j\vec E^ {extra} dV - \int {| \vec j |
         ^2\over \lambda} dV, \end {equation}
according to which it follows that the increment of the energy of an
electromagnetic field is equal to an excess of the work of the
extraneous electromotive forces over the heat generation. This is the
law of the conservation of energy in electrodynamics including the
motionless bodies \cite {tamm}.

In this case the proof is strict since the energy of the material
bodies of an electromagnetic origin is absent or in the case of charged
bodies is constant.

\section {Introduction of particles in electrodynamics}

We have shown, that expression (5), does not
describe the law of conservation of energy in electrodynamics of
Maxwell-Lorentz. In the present section particular examples
are considered allowing directly i.e. without basing on expression
(5) to illustrate inconsistency of the equations of motion of
particles and field in electrodynamics of Maxwell-Lorentz or, at least,
to impose certain restrictions on a possible type of fields of
non-electromagnetic origins. In this case we shall proceed from the
existence of the law of conservation energy for the electromagnetic
field and particles in unknown for the present form which is differ
from (5).

\subsection {Example 3}

Two identical charged particles are brought closer to each other with
equal constant velocities by extraneous forces along an axis $x$ up to
a distance $a_1$.  Then the extraneous forces are switched off and
particles being decelerated continue to be brought closer by inertia
until they will stop on a distance $a_2 < a_1$. After the stop of
particles the extraneous forces are switched on again to keep particles
in the state of rest.

Show that contrary to the law of conservation of energy\footnote
{\normalsize As the repulsive forces between moving particles $eE_
{||} ^{mov} $ are weakened $\gamma ^2$ times in comparison with a
static case then on a principle of bringing closer of particles to each
other with relativistic velocities and subsequent separation them
under non-relativistic velocities one could construct the perpetum
mobile.} the energy of the considered system in the final state is higher
than the energy spent by extraneous forces on acceleration of particles
and their further bringing closer.

The idea of this example is the next. The electrical field of a
uniformly moving particle is flattened in the direction of motion
such a way that on the axis of motion at a distance $a$ its value
$E_{||}$ is $\gamma ^2$ times less then the electrostatic
field of a charge at rest being at the same distance from the
observation point \cite {landau}-\cite {sokol}
         \begin {equation}  
          E_ {||} ^ {mov} = {1\over \gamma ^2} E_ {||} ^ {rest} =
         {e^2\over \gamma ^2 a^2} \end {equation}

To transform to particles the initial velocity and bringing closer them
up to a distance $a_1$ the extraneous forces must transform them the
energy
                \begin {equation} 
         \varepsilon _1 = 2 mc^2 (\gamma -1) + {e^2\over \gamma ^2 a_1}
         + \varepsilon _ {1rad} ^ {em}, \end {equation}
where $\varepsilon _ {rad} ^ {em} $ is the energy radiated by particles
in the process of acceleration.

The energy of system in the final state is equal
           \begin {equation} 
         \varepsilon _2 = {e^2\over a_2} +
         \varepsilon _ {1rad} ^ {em} + \varepsilon _ {2rad} ^ {em},
          \end {equation}
where $\varepsilon _ {2rad} ^ {em} $ is the energy of radiation emitted
by particles at the process of deceleration.

The value
           \begin {equation} 
         \varepsilon _2 - \varepsilon _1
        = {e^2\over a_2} - {e^2\over \gamma ^2 a_1} +
        \varepsilon _ {2rad} ^ {em} - 2mc^2 (\gamma -1),
          \end {equation}
which is equal to a difference between the energy transformed by
charged particles to extraneous forces at separation of these particles
and energy transformed by extraneous forces to particles to bring
closer the particles.

It follows from (13) and conditions $\varepsilon _ {2rad} ^ {em} >
0$, $a_2 < a_1$ that in this case there is an excess in energy
$\Delta \varepsilon = \varepsilon _2 - \varepsilon _1> 0$
which is differ from zero at least at
               \begin {equation} 
         A_1 < r_e {\gamma + 1\over 2\gamma ^2}, \end {equation}
where $r_e = e^2/mc^2$ is the classical radius of a particle.

The function $\Delta \varepsilon (a_1) $ is obviously analytic
one\footnote {\normalsize Remind that if two analytical functions
coincide within some range of variables then they are identically equal
through the whole range of these functions.}. It means that $\Delta
\varepsilon \ne 0$ at arbitrary $a_1$.

In this case if we start from the low of conservation of energy of the
unknown for the present form which is differ from (5) then we are
forced to impose restrictions on the range of applicability of
the Maxwell equations and in particular on the Coulomb's low
at small distances $a \leq r_e$. It is possible also to conclude that
the fields of a non-electromagnetic origin can not be short-range one.
After a stop of particles these fields lead to an attraction between
them on distances $a \leq r_e$ with forces comparable with
electrostatic repulsive forces. It means that in this case the
classical electrodynamics cannot be built without the account of forces
of a non-electromagnetic origin or that it is necessary all at once to
build the unified theory of fields where the fields of
non-electromagnetic origin on the same level with electromagnetic
fields will be described by definite equations\footnote {\normalsize It
is possible to accept that in the equations of motion of charged
particles and further in expressions (2),(5) a mass $m_i$ and an energy
$\varepsilon _i$ have a non-electromagnetic origin. In this case it
would be possible to interpret the inertial coefficient automatically
arising in the equation of a motion of particles as a mass of an
electromagnetic origin and the energy of particles in (5) would be
composed from a part of an energy of electromagnetic origins (first
term) and from an energy $\varepsilon _i$ of non-electromagnetic
origin. However such approach would be an acceptable way of introduction
of fields of non-electromagnetic origin in the electrodynamics of
Maxwell-Lorentz (the way based on a postulated law of
conservation of energy) but would not be the proof of existence of
this law in the framework of the electrodynamics.}.

It is possible to assume also, that the size of particles should be
comparable with the size $r_e$. Then the position of particles
can not be determined by three coordinates and at the description of
the particle's motion which draw closer to one another up to a distance
$a \sim r_e$ it is necessary to take into account their internal
structure.

\subsection {Example 4}

Show that the energy of an electromagnetic field radiated by a particle
moving along a line of forces of a homogeneous electrical field of a
capacitor under definite value of the field strength can be higher than
the energy transformed by the field of the capacitor to the particle
and find the value of the field.

The energy of an electromagnetic field radiated by a particle is equal
           \begin {equation} 
         \varepsilon _ {rad} ^ {em} = {2e^4E^2\over 3m^2c^3} \Delta t,
          \end {equation}
where $\Delta t = (cm/eE) \sqrt {(eEl/mc^2 + 1) ^2 - 1} $ is the time
of the particle  motion in the capacitor of the length $l$ [1-4].

The energy transferred by the electrical field to the particle is equal
         \begin {equation} 
         \Delta \varepsilon = eEl.  \end{equation}
It should follow from the law of conservation of energy that this
work is equal to a sum of the emitted energy $\varepsilon _ {rad} ^
{em} $ and the kinetic energy of a particle $mc^2 (\gamma -1) $ or
         \begin{equation} 
         \Delta \varepsilon = \varepsilon _ {rad} ^
         {em} + mc^2 (\gamma - 1).  \end {equation}

But the relation (17) cannot be valid at any fields $E$ in view of the
fact that $\Delta t> l/c$, $mc^2 (\gamma -1) > 0$, $\varepsilon _ {rad}
^ {em} \sim E^2$, $\Delta \varepsilon \sim E$. It is violated at the
fields
            \begin {equation} 
            E> {3e\over r_e^2}.  \end{equation}

The energy of the electromagnetic radiation emitted by the particle in
the capacitor at such field strengths will be higher than the energy
that one needs to move it from one capacitor plate to another. This
result contradicts to the law of the conservation of energy.

It is possible to eliminate this contradiction if to assume that
particles in the classical electrodynamics have the final dimensions.
In this case in (15) there will appear the coherence factor of
the radiation which is decreased when the value $E$ is increased. To
assume that the radiation reaction forth appear in the homogeneous
electrical field for charged particles, which is increased with
increasing of the value $E$ is impossible without the conflict with the
equations of motion \cite {landau},\cite{ginzb}.

             \section {Conclusion}

In electrodynamics there are many "open" or "perpetual" problems such
as the problem of the self-energy and momentum of particles, the nature
of the particle's mass, the problem of the runaway solutions.
There is a spectrum of opinions concerning the importance and the ways
of a finding of the answers on these questions. A number of
these opinions we shall give below.

Unfortunately the efforts of the majority of the authors are directed
more often to avoid similar questions than to solve them. In addition
they base themselves on the laws of conservation ostensibly following
from the electrodynamics in the most general case and presenting
electrodinamics as
the consistent theory. In such stating the arising questions do not
have a physical subject of principle and the difficulties in their
solution are on the whole in the field of the mathematicians.

It is shown in the present work that a relation (5) does not express
the law of conservation of energy in electrodynamics of
Maxwell-Lorentz.  The error in the treatment of this expression is the
consequence of insufficiently precise definitions of the basic concepts
of the theory and its logically inconsistent construction. It is
shown also that this electrodynamics is incompatible with the concept
of pointlike particles even in that case, when a correct transformation
properties are attached to them.

It follows, in particular, that in the process of any discussion of the
existing difficulties of the theories it is impossible to refer on the
law of conservation of energy in electrodynamics in the form, which was
done, for example, in the textbooks
\cite {landau},\cite {ginzb},\cite{markov}.

We hope that this work will permit to produce a more comprehensive 
analysis and to stimulate the further development of the foundations of 
the classical and quantum electrodynamics.

\newpage

             \section {Appendix}

The electromagnetic field in vacuum is described by the Maxwell
equations

          \begin {equation} 
           rot \vec E = - {\partial \vec H\over \partial t},
          \end {equation}
          \begin {equation} 
           div \vec H =0,
           \end {equation}
           \begin {equation} 
            rot \vec H = {4\pi\over c} \vec j + {1\over c} {\partial \vec E\over
           \partial t}
           \end {equation}
           \begin {equation} 
            div \vec E = 4\pi \rho
           \end {equation}

The typical proof of the law of conservation of energy in
electrodynamics is derived according the following scheme \cite
{landau}.

Let us multiply both parts of the equation (19) on $\vec H$ and both
parts of the equation (21) on $\vec E$ and subtract the received
equations term by term

           \begin {equation} 
          {1\over c} (\vec E {\partial \vec E\over \partial t}) +
          {1\over c} (\vec H {\partial \vec H\over \partial t}) =
          - {4\pi\over c} \vec j - (\vec H rot \vec E - \vec E rot \vec
           H). \end {equation}

Using the known formula of the vector analysis $div [\vec a\vec b] =
\vec b rot\,\vec a - \vec a rot\,\vec b$ we rewrite this relation in
the form
            \begin {equation} 
          {\partial \over \partial t} {| \vec E | ^2 + | \vec H | ^2\over 8\pi}
            = - \vec j \vec E - div \vec S, \end {equation}
where $\vec S= (c/4\pi) [\vec E\vec H] $ is the Pointing vector.

Let us integrate (24) through some volume and apply the Gauss theorem
to the second term from the right. Then we shall receive the equation
              \begin {equation} 
              \int ({\partial \over \partial t} {| \vec E | ^2 + | \vec
              H | ^2\over 8\pi} + \vec j\vec E) dV = - \oint \vec S
              d\vec f, \end {equation}
where $d\vec f = \vec n d\, f$ is a vector of an element of a surface
determined by the area $d\, f$ and unit vector $\vec n$ normal to the
surface and directed outside of the closed volume. If the system
consists of charged particles then, according to (2), the integral
$\int \vec j \vec E dV$ is written down in the form of a sum
corresponding to all particles of system of a form $\sum e\vec
V_i\vec E (\vec r_i) = \sum d\varepsilon _i/dt$.  In this case (25)
is transformed in
              \begin {equation} 
             {\partial \over \partial t} (\int {| \vec E | ^2 + | \vec
              H | ^2\over 8\pi} dV + \sum _i \varepsilon _i) = - \oint
              \vec S d\vec f.  \end {equation}

The value {\Large $\oint $} $\vec S d\vec f$ is a flow of the energy of
the electromagnetic field through a surface limiting the volume. If the
integration is made through the total volume of the space i.e. if a
surface of integration is withdrawn to infinity then the surface
integral is disappeared (all fields and particles remain in space and
do not go outside of the limits of the surface of integration) \cite
{landau}. It follows the expression (5).

Notice that at the derivation of the expression (5) the expression (2)
was used which is derived in the frameworks of relativistic mechanics 
independently on any introduction of the Maxwell equations and
consequently concepts of the density of the energy of the
electromagnetic field, energy and mass of particles of electromagnetic
origins \cite {landau}. It is natural that in this case the mass and
the energy of particles continue to be considered as some mechanical
characteristics of particles and the fact is lost that with the charged
particle the led self-fields are connected extending to the whole
space, having an energy of an electromagnetic origin included in the
first term of the expression (5), causing at acceleration the inertial
and radiation reaction forces.

After introduction of the Maxwell equations and postulate (4) it would
follow to postulate the law of conservation of energy for particles
and field and being based on this base to search for ways of
introduction of particles to classical electrodynamics just as it
was done by Abraham and Lorentz \cite {jack} or to search the equations
for fields of a non-electromagnetic origin and also the other ways of
the solution of the problem.


\newpage 
\vskip 5cm \empty
\vspace * {1cm}

\section {Spectrum of opinions}

\vskip 5mm

For some reason, the controversies which relate to the problems in
understanding the basic principles of the theory of relativity
and of quantum mechanics usually take a particularly acute form. they
often "become personal," with mutual accusations of ignorance, etc.\\

\vskip 2mm \empty \hskip 2cm E.~L.~Feinberg, Sov. Phys. Uspekhi, Vol.
18, No 8, (1976) \vskip 20mm

... In the field theory the authors come, for example, to a conclusion that
"elementary particles cannot have of the final dimensions and should be
considered as geometrical points."  Such conclusion is hardly
comprehended physically and philosophically and in any case requires
physical and philosophical comment especially in the textbooks.
Otherwise the reader of the textbook, a physicist or philosopher may
do the conclusion that for Landau and Lifshitz "space is not the form
of existence of matter" or that the elementary particles does not exist
as there is nothing in a mathematical point. .....

Our duty is to achieve the state when in each general and special
guide of physics - I mean in particular printed textbooks on physics -
the philosophical base will find its clear expression. For the reader
it is necessary the comprehensive organic penetration of correct
philosophy into a specific matter. \\

\vskip 2mm \empty \hskip 1cm S.~I.~Vavilov, Philosophskie voprosy
sovremennoi physiki, 
Izdat. AN, SSSR, Moscow, 1952

\vskip 20mm
\empty \vspace * {0.5cm}
Before it has become clear many theorists (Helmholtz, Hertz,
Zommerfeld and other) have written multivolume, not to tell monstrous,
works about electromagnetic self-energy of the hard electron ....
Nowadays all these efforts seem to be vain; the quantum theory has
replaced this point of view and now the tendency consists rather in
avoiding of a problem of self-energy, than to solve it.  But there
will come a day when it will become central one again. \\

\vskip 2mm \empty \hskip 1cm
M.Born, Lecture read in Bern 16.7.1955 at an International conference
devoted to 50 years\\
\vskip - 5mm \empty \hskip 1.5cm
of the theory of relativity (Naturwiss. Rundschaw), 1956 (from A.~A.Typkin,
Principle \\
\vskip - 5mm \empty \hskip 5cm
of relativity, Atomizdat, Moscow, 1975, p.23)

\newpage

\vskip 2cm
There is one strange moment in the previous consideration. Classical
electrodynamics is a relativistic covariant theory. That is why it is
possible to wait that correct calculation of any value the requirements
of the Lorentz covariance will not be violated. Nevertheless in the
Abraham-Lorentz model, apparently, such violation is available. The
non-covariant electromagnetic part of self-energy and momentum of a
charged particle is counterbalanced in this model by the non-covariant
part made dependent on Poincare stresses such a way that the result
has a relativistic covariance. Certainly it is possible to say that
since for a stable state of a limited distribution of a charge the
retained forces of a non-electromagnetic origin and corresponding
fields are necessary then only total forces have a physical sense.
A question is nevertheless natural - Is it possible to determine the
purely electromagnetic part of self-energy and momentum such a way that
it has the relativistic covariance? Such determination would have not
only aesthetic value but would separate at least formally the question
of stability from a question of Lorentz invariance.

\vskip 2mm \empty \hskip 2cm
Dzh. Dzhekson, {\it Classicheskaya electrodynamica,} World, Moscow,
1965, p.651; \\
\vskip - 5mm \empty \hskip 2cm
see also: J.~D.~Jackson, {\it Classical Electrodynamics,}
John Wiley $\&$. Sons, 1962 \\

\empty \vspace * {0.1cm}
\vskip 1cm
... a completely satisfactory treatment of the reactive effects of
radiation does not exist. The difficulties presented by this problem
touch one of the most fundamental aspects of physics, the nature of an
elementary particle. .... one might hope that the transition from
classical to quantum-mechanical treatments would remove the
difficulties. While there is still hope that this may eventually occur,
the present quantum-mechanical discussions are best with even more
elaborate troubles than classical ones. It is one of the triumphs of
comparatively recent years ($\sim$1948-1950) that the concepts of
 Lorentz covariance and gauge invariance were exploited sufficiently
cleverly to circumvent these difficulties in quantum electrodynamics.
... From a fundamental point of view, however, the difficulties still
remain.

\vskip 2mm
\empty \hskip 0.5cm
J.~D.~Jackson, {\it Classical Electrodynamics,} John Wiley $\&$. Sons,
1975, p.781 \\

\vskip 1.5cm
In physics there are many "perpetual problems" the discussion of
which contains for decades in the scientific literature let alone in
textbooks.

The problem of radiation from a uniformly accelerated charge and most
of other "perpetual problems " are undoubtedly of no major
significance and this is precisely why they have remained
insufficiently well explained for so long time. On the other hand,
however, neglect of such methodical types of problems sometimes incurs
vengeance.

\vskip 2mm \empty \hskip 2cm
V.~L.Ginzburg, Soviet Physics Uspekhi, v.12, No 4, p.565 (1970)

\vskip 15mm \empty \hskip 1cm
See also: R.~P.~Feynman, R.~B.~Leyghton, and M.~Sands, The Feynman
Lectures \\
\vskip - 5mm \empty \hskip 3cm
on Physics, v.2, Addison-Wesley, Reading, Mass. (1963).

\newpage
\empty \vspace * {15cm} 
\begin{center} Signed to print April 25, 1997

Order No 145. 50 copies printed. P.l. 1

--------------------------------------

Printed in RIIS FIAN.

Moscow, W-333, Leninsky prospect, 53.
\end{center}
\footskip=100mm
\newpage}
\end {document}